\documentclass[aps,pre,twocolumn,superscriptaddress,showpacs]{revtex4-1}
\usepackage[dvips]{graphicx}
\usepackage{amsmath}
\begin{document}


\title{Synchronization in the random field Kuramoto model on complex networks}


\author{M. A. Lopes}
\affiliation{Department of Physics $\&$ I3N, University of Aveiro, 3810-193 Aveiro, Portugal}
\affiliation{College of Engineering, Mathematics and Physical Sciences, University of Exeter, UK}
\affiliation{Wellcome Trust Centre for Biomedical Modelling and Analysis, University of Exeter, UK}
\author{E. M. Lopes}
\affiliation{Department of Physics $\&$ I3N, University of Aveiro, 3810-193 Aveiro, Portugal}
\author{S. Yoon}
\affiliation{Department of Physics $\&$ I3N, University of Aveiro, 3810-193 Aveiro, Portugal}
\author{J. F. F. Mendes}
\affiliation{Department of Physics $\&$ I3N, University of Aveiro, 3810-193 Aveiro, Portugal}
\author{A. V. Goltsev}\email{goltsev@ua.pt}
\affiliation{Department of Physics $\&$ I3N, University of Aveiro, 3810-193 Aveiro, Portugal}
\affiliation{A.F. Ioffe Physico-Technical Institue, 194021 St. Petersburg, Russia}




\begin{abstract}
We study the impact of random pinning fields on the emergence of synchrony in the Kuramoto model on complete graphs and uncorrelated random complex networks. We consider random fields  with uniformly distributed directions and homogeneous and heterogeneous (Gaussian) field magnitude distribution. In our analysis we apply the Ott-Antonsen method and the annealed-network approximation to find the critical behavior of the order parameter. In the case of homogeneous fields, we find a tricritical point above which a second-order phase transition gives place to a first-order phase transition when the network is either fully connected, or scale-free with the degree exponent $\gamma>5$. Interestingly, for scale-free networks with $2<\gamma \leq 5$, the phase transition is of second-order at any field magnitude, except for degree distributions with $\gamma=3$ when the transition is  of infinite order at $K_c=0$ independently on the random fields. Contrarily to the Ising model, even strong Gaussian random fields do not suppress the second-order phase transition in both complete graphs and scale-free networks though the fields increase the critical coupling for $\gamma > 3$. Our simulations support these analytical results.
\end{abstract}

\maketitle

\section{Introduction}
\label{introduction}
The Kuramoto model has been the paradigmatic model to study synchronization phenomena in a multitude of fields, from physics to sociology \cite{Strogatz_2000,Pikovsky_2003,Acebron_2005,rodrigues2016kuramoto}. The model describes the dynamics of interacting phase oscillators. Depending on how strong the coupling is, the oscillators may be in a disordered state, or in a ordered state, i.e., oscillating synchronously with a certain frequency, despite of the heterogeneity in their natural frequencies. Kuramoto showed that, for all-to-all interaction, synchronization emerges as a result of a second-order phase transition \cite{Kuramoto_1975,Kuramoto_1984}. Since then, most works on the Kuramoto model described continuous phase transitions that occur for different generalizations of the model. First-order phase transitions (also called explosive synchronization in this context) were found as well for other generalizations of the model, namely with time delays \cite{Yeung_1999},  frequency-degree correlations \cite{gomez2011explosive,bruno_coutinho}, and frequency weighted coupling 
\cite{zhang2013explosive} (see reviews \cite{Acebron_2005,rodrigues2016kuramoto} for other examples).

Our previous findings on the critical behavior of the order parameter, relaxation rate, and susceptibility of the Kuramoto model on uncorrelated random complex networks demonstrated that this model has the same critical exponents as the Ising model, and therefore it should belong to the same class of universality \cite{Yoon_2015}. This was shown for the Kuramoto model in the presence of a uniform external field. More generally, one should also analyze the impact of random fields (random pinning) on the critical behavior as in the random field Ising model on complex networks \cite{dhar1997zero,Dorogovtsev_2008,Dorogovtsev_2008arxiv,Lee_2006}.
The consideration of random fields may also be crucial to understand how disorder affects the emergence of synchrony in real systems such as cortical oscillations or the circadian clock in the brain \cite{liu1997cellular,breakspear2010generative}. The random field Kuramoto model with all-to-all coupling was first studied in the case when random fields have the same magnitude and their directions are uniformly distributed \cite{strogatz1989collective}.
In this case,  a sufficiently strong random pinning results in a first-order phase transition.
On the other hand, if the magnitudes are random but the fields have the same direction, then strong pinning can suppress the synchronization \cite{arenas1994exact}.

In this paper we present the first analytical and numerical treatment of the Kuramoto model with heterogeneous random fields in random complex networks. We demonstrate that the network topology and the random field heterogeneity impact strongly both on the critical coupling and on the kind of the synchronization phase transition, which can be of both first- and second-order, or infinite-order.

The paper is organized as follows. In Sec. \ref{model} we present the random field Kuramoto model on complex networks using the annealed network approximation. We use the Ott-Antonsen method to reduce the problem to a single differential equation for the order-parameter. In Sec. \ref{critical} we solve the model for a complete graph and scale-free networks
in the presence of homogeneous and heterogeneous (Gaussian) random fields. Our results are summarized and discussed in Sec. \ref{conclusion}.

\section{Random field Kuramoto model
}
\label{model}
\subsection{Annealed network approximation}
The original Kuramoto model describes the evolution of $N$ phase oscillators according to
the following
equations,
\begin{equation}
\frac{d \theta_i}{dt}=\omega_i + \frac{K}{N}\sum_{j=1}^N\sin(\theta_j-\theta_i),
\end{equation}
where $\theta_i$ is the phase of oscillator $i$, $\omega_i$ is its natural frequency, and $K$ is the coupling constant.
The oscillators' natural frequencies are heterogeneous and follow a probability density function $g(\omega)$. Despite this heterogeneity, the oscillators become synchronized for sufficiently large $K > K_c$. The order parameter is defined as the fraction of synchronized oscillators,
\begin{equation}
z=re^{i\psi}\equiv \frac{1}{N}\sum_{j=1}^Ne^{i\theta_j},
\end{equation}
where the magnitude $r$ characterizes the phase coherence, and $\psi$ is the collective phase. $r$ varies between $0$ and $1$, which corresponds to total disorder and complete synchronization, respectively.

A natural generalization of the model is to consider interacting oscillators on a complex network described by an adjacency matrix $a_{ji}$ ($a_{ji}=1$ if $j$ is connected to $i$, and $a_{ji}=0$ otherwise),
\begin{equation}
\frac{d \theta_i}{dt}=\omega_i + K\sum_{j=1}^Na_{ji}\sin(\theta_j-\theta_i).
\end{equation}
In order to analyze this equation, one can use the so-called 'annealed network' approximation \cite{Dorogovtsev_2008,Bianconi_2002,bruno_coutinho} which replaces a random complex network
by a weighted complete graph in which edge weights are the probabilities of connections between nodes in the original graph,
\begin{equation}
\langle a_{ji} \rangle = \frac{q_j q_i}{N\langle q \rangle},
\label{eq:annealed}
\end{equation}
where $q_i$ is the degree of node $i$ and $\langle q \rangle$ is the mean degree.
Thus the contribution of each oscillator is weighted by its degree,
\begin{equation}
\frac{d \theta_i}{dt}=\omega_i + \frac{Kq_i}{N\langle q \rangle}\sum_{j=1}^Nq_j\sin(\theta_j-\theta_i),
\label{eq:kuramoto_annealed}
\end{equation}
and the order parameter is redefined as
\begin{equation}
z=re^{i\psi}\equiv \frac{1}{N\langle q \rangle}\sum_{j=1}^Nq_je^{i\theta_j}.
\label{eq:order_parameter}
\end{equation}
Using (\ref{eq:order_parameter}) one can write Eq.~(\ref{eq:kuramoto_annealed}) as follows:
\begin{equation}
\frac{d \theta_i}{dt}=\omega_i + Kq_ir\sin(\theta_i-\psi).
\end{equation}

\subsection{Random fields}
Using the similarity between the XY-model and the Kuramoto model, one can introduce local fields $\vec{h}_i$ acting on the phase oscillators. Each field $\vec{h}_i=(h_x(i),h_y(i))=h_i(\cos\phi_i,\sin\phi_i)$ is characterized by a magnitude $h_i=(h^{2}_x(i) +h^{2}_y(i))^{1/2}$ and an angle $\phi_i$. The interaction energy between a field $\vec{h}_i$ and an oscillator $i$ is $E=-\vec{h}_i\cdot\vec{n}_i$, where $\vec{n}_i(\theta)=(\cos\theta_i,\sin\theta_i)$ is the unit vector characterizing oscillator $i$. Consequently, the force of the local field upon the oscillator is $-\partial E / \partial \theta_i=h_i\sin(\phi_i-\theta_i)$. The Kuramoto model on complex networks in the presence of local fields is then described by the following equations \cite{Shinomoto_1986,Yoon_2015}:
\begin{equation}
\frac{d \theta_i}{dt}=\omega_i + Kq_ir\sin(\theta_i-\psi)+h_i\sin(\phi_i-\theta_i).
\label{eq:kuramoto_field}
\end{equation}
The additional term produces the pinning effect meaning that it tries to force each oscillator to be stuck at a random angle. Therefore, it favors a static disordered state.
Note that we consider the case when the random fields rotate with frequency equal to the group velocity $\Omega$. So, in the rotating frame, the random fields are static.
As we already mentioned in the introduction, this pinning term was studied within the Kuramoto model on a complete graph in some particular cases \cite{strogatz1989collective,arenas1994exact}.
Here, we consider the case when the phases of the local fields are uniformly distributed in $[0,2\pi)$.
The probability density distribution of local random fields $\vec{h}_i$ is $G(h_x(i),h_y(i))=f(\phi_i )G(h_i)$ where $f(\phi_i)=1/(2\pi)$ is the uniform distribution of the local fields' phases and $G(h_i)$ is the probability density distribution of the fields' magnitude. The normalization condition is
\begin{equation}
\int \int G(h_x,h_y) dh_x dh_y =\int_{0}^{2 \pi}  f(\phi)d\phi \int_{0}^{\infty} G(h) h dh  =1.
\label{eq:h-normalization}
\end{equation}
Regarding to the field magnitude, we study two cases. First, all local fields have the same magnitude, i.e., $h_i=h$. In this case we have
\begin{equation}
G(h_i)=\frac{1}{h} \delta(h_i-h),
\label{eq:homogeneous}
\end{equation}
Second, the entries $h_x(i)$ and $h_y(i)$ are Gaussian distributed, then
\begin{equation}
G(h_i)=\frac{1}{\sigma^2}\exp \Bigl[-\frac{h_x^2(i)+h_y^2(i)}{2\sigma^2}\Bigr].
\label{eq:gaussian}
\end{equation}

\subsection{Ott-Antonsen method}
We use the Ott-Antonsen method \cite{Ott_2008,Ott_2009} to find a set of differential equations for the time evolution of the order parameter $z$ in Eq. (\ref{eq:order_parameter}). We follow the same approach as in \cite{Yoon_2015}. In the limit $N\to\infty$, we define the oscillator density $F(\theta,\omega,q,h,\phi,t)$ that satisfies the normalization conditions,
\begin{equation}
\int_0^{2\pi}\int_0^\infty\int_1^\infty\int_0^{2\pi} F(\theta,\omega,q,h,\phi,t)hd\theta dq dh d\phi=g(\omega),
\end{equation}
\begin{equation}
\int_0^{2\pi}\int_0^\infty\int_{-\infty}^\infty\int_0^{2\pi} F(\theta,\omega,q,h,\phi,t)hd\theta d\omega dh d\phi=P(q),
\end{equation}
\begin{equation}
\int_0^{2\pi}\int_{-\infty}^\infty\int_1^\infty\int_0^{2\pi} F(\theta,\omega,q,h,\phi,t)d\theta dq d\omega d\phi= hG(h),
\end{equation}
and
\begin{equation}
\int_{-\infty}^{\infty}\int_0^\infty\int_1^\infty\int_0^{2\pi} F(\theta,\omega,q,h,\phi,t)hd\theta dq dh d\omega=f(\phi)
\end{equation}
Note that here we replaced the sum over the degrees $q$ by the integration.

The oscillator density obeys the conservation law,
\begin{equation}
\frac{\partial F}{\partial t}+\frac{\partial [\nu F]}{\partial \theta}=0,
\label{eq:conservation}
\end{equation}
where $\nu$ is the velocity
that drives the dynamics of the oscillators,
\begin{eqnarray}
\nu&=&\omega + Kqr\sin(\theta-\psi)+h\sin(\phi-\theta)\\
&=&\omega+\frac{Kq}{2i}(ze^{-i\theta}-z^*e^{i\theta})+\frac{h}{2i}(e^{i(\phi-\theta)}-e^{-i(\phi-\theta)})\nonumber
\end{eqnarray}

Following the Ott-Antonsen method \cite{Ott_2008,Ott_2009}, we look for a solution in the form
\begin{equation}
F=\frac{g(\omega)P(q)G(h)f(\phi)}{2\pi}(1+F_++F_-),
\label{eq:F}
\end{equation}
where
\begin{equation}
F_+=\sum_{n=1}^\infty \alpha^n(\omega,q,h,\phi,t)e^{in\theta}
\end{equation}
and $F_-=F_+^*$. Substituting Eq.~(\ref{eq:F}) into Eq.~(\ref{eq:conservation}), we obtain
\begin{equation}
\dot{\alpha}+i\omega\alpha+\frac{Kq}{2}(z\alpha^2-z^*)+\frac{h}{2}(\alpha^2e^{i\phi}-e^{-i\phi})=0,
\end{equation}
where $z$
can be written as
\begin{eqnarray}
z(t)&=&\int_0^{2\pi}f(\phi)d\phi\int_0^\infty G(h)hdh \int_1^\infty \frac{qP(q)}{\langle q \rangle}dq\times  \nonumber \\
& &\int_{-\infty}^\infty g(w)\alpha^*(\omega,q,h,\phi,t)d\omega .
\label{eq:z}
\end{eqnarray}
As in Ref. \cite{Yoon_2015}, we look for the stationary solution ($\dot{\alpha}=0$) at which the phase coherence $r$ is constant.
We set $\psi=0$ so that $z=r$. With these conditions, we find the steady state solution
\begin{equation}
\alpha_0=
\begin{cases}
\frac{A\sqrt{C}-\omega B-i(\omega A+B\sqrt{C})}{A^2+B^2}, &\text{if } |\omega| \le \sqrt{A^2+B^2},\\
\frac{B\sqrt{-C}-\omega B+i(A\sqrt{-C}-\omega A)}{A^2+B^2}, & \text{if } |\omega| > \sqrt{A^2+B^2},\\
\end{cases}
\end{equation}
where $A=Kqr+h\cos\phi$, $B=h\sin\phi$ and $C=A^2+B^2-\omega^2$. Finally, we take the real part of Eq.~(\ref{eq:z}). In the case of a Lorentz distribution of natural frequencies, $g(\omega)=\Delta / \pi (\omega^2+\Delta^2)$, with $\Delta=1$ as frequency unit, and an uniform distribution of local fields phases, $f(\phi)=1/2\pi$, we obtain a nonlinear equation determining $r$ as a function of $K$, degree and random field distributions, $P(q)$ and $G(h)$,
\begin{eqnarray}
\label{eq:r}
r&=&\frac{1}{2\pi}\int_0^{2\pi}d\phi \int_0^\infty G(h)hdh \int_1^\infty \frac{qP(q)}{\langle q \rangle}\times \\
& &\frac{Kqr+h\cos\phi}{\sqrt{(Kqr+h\cos\phi)^2+h^2\sin^2\phi+1}+1} dq . \nonumber
\end{eqnarray}

\section{Critical behavior of the order parameter}
\label{critical}
In this section we find the critical behavior of the order parameter for the Kuramoto model on a complete graph and scale-free networks in the presence of either homogeneous random fields, Eq.~(\ref{eq:homogeneous}), or Gaussian random fields, Eq.~(\ref{eq:gaussian}). In the case of a complete graph, the degree distribution is $P(q)=\delta(q-N+1)$ in Eq.~(\ref{eq:r}) since all nodes
have the same degree $q=N-1$, whereas for a scale-free graph we use $P(q)=Cq^{-\gamma}$, where $C$ is the normalization constant.
Note that the annealed network approximation, Eq.~(\ref{eq:annealed}), requires a finite first-moment, thus limiting our analysis to $\gamma>2$.
We take the Taylor expansion of the right hand side (RHS) of Eq.~(\ref{eq:r}) with respect to the order parameter $r$.
By denoting the RHS as $F(r)$ and taking into account that even terms are zero, we obtain
\begin{equation}
r=F'(0)r+F'''(0)\frac{r^3}{3!}+O(r^5).
\label{eq:expansion}
\end{equation}
In the leading order in $r$, the condition $F'(0)=1$ defines the critical coupling $K_c$.
For a complete graph and scale-free networks with $\gamma>5$, using the next order in $r$, we find \begin{equation}
r=\sqrt{\frac{6[1-F'(0)]}{F'''(0)}},
\label{eq:r_meanf}
\end{equation}
corresponding to the mean-field exponent $\beta=1/2$.

In the case of scale-free networks, $F'(0)\propto \int_1^\infty P(q) q^2 dq = \langle q^2 \rangle$, and $F'''(0)\propto \langle q^4 \rangle$.
Consequently, $F'''(0)$ diverges at $\gamma \leq 5$ and this case demands a more careful analysis of Eq.~(\ref{eq:r}), which we present below.

\subsection{Complete graph}
For a complete graph in the presence of homogeneous fields with a constant magnitude $|\vec{h}_i|=h$,
we find explicitly the field dependence of the critical coupling
$K_c$ and the order parameter $r$,
\begin{eqnarray}
&K_c=2\sqrt{h^2+1},
\label{eq:Kc_complete homogeneous}
\\
&r=\frac{K_c^2}{2}\sqrt{\frac{2}{K^3(2-h^2)}}(K-K_c)^{1/2}.
\label{eq:complete_homogeneous}
\end{eqnarray}
At $h=0$, we recover the classical expressions for $K_c$ and $r$ \cite{Kuramoto_1975}. According to Eq. (\ref{eq:Kc_complete homogeneous}), the larger the magnitude $h$ of the random fields, the larger has to be the coupling $K$ in order to synchronize the oscillators.

\begin{figure}
\includegraphics[width=0.4\textwidth]{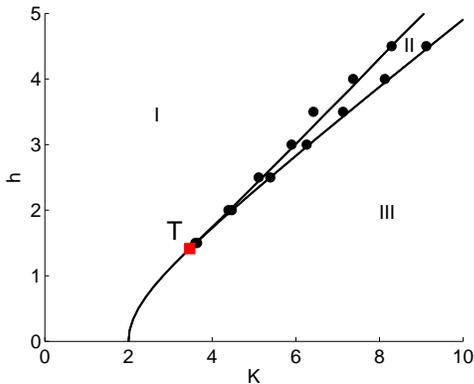}
\caption{$K$-$h$ phase diagram of the Kuramoto model on a complete graph in the presence of homogeneous random fields of magnitude $h$. The solid line corresponds to the numerical solution of Eq.~(\ref{eq:r}) and the points to our simulations.
There are three regions: (I) asynchronous state, (II) region of hysteresis, and (III) partially synchronous state. $T=(K_c,h_c)=(2\sqrt{3},\sqrt{2})$ is a tricritical point below which there is a second-order phase transition, whereas above it there is a first-order phase transition. The networks were generated using the static model ($N=10^4$, and $\langle q \rangle=10$) \cite{Goh_2001,Lee_2004}. The differential equations for each oscillator were solved using a fourth-order Runge-Kutta method with time step $\Delta t=0.01$.
\label{fig:hk_complete}}
\end{figure}

However, if $h>\sqrt{2}$, then the right hand side in Eq.~(\ref{eq:complete_homogeneous}) becomes imaginary. It means that the approximation $r\ll1$ is no longer correct. Indeed, a numerical solution of Eq.~(\ref{eq:r}) shows that at $h>\sqrt{2}$ the synchronization transition becomes  discontinuous. In Fig.~\ref{fig:hk_complete} we show this solution and compare it with simulations of the model for $N=10^4$. One can see that in the $K$-$h$ plane there is a tricritical point $(K_c,h_c)=(2\sqrt{3},\sqrt{2})$ where a second-order phase transition meets a first-order phase transition. Furthermore, the region of hysteresis becomes larger as the random field magnitude $h$ is increased.

If instead of homogeneous random fields, we consider random fields with Gaussian magnitudes, we find
\begin{eqnarray}
&K_c=\frac{1}{\int_0^\infty h \ G(h)(2\sqrt{h^2+1})^{-1}dh},
\label{eq:f}
\\
&r=\sqrt{\frac{1}{K_c K^3 \int_0^\infty \ h \ G(h) \ v(h)dh}} \ (K-K_c)^{1/2},
\label{eq:complete_gaussian}
\end{eqnarray}
where
\begin{equation}
v(h)=\frac{2-h^2}{16(h^2+1)^{5/2}},
\end{equation}
and $G(h)$ corresponds to Eq.~(\ref{eq:gaussian}) (if we use Eq.~(\ref{eq:homogeneous}), we get Eqs.~(\ref{eq:Kc_complete homogeneous}) and (\ref{eq:complete_homogeneous}), respectively).
In this case, Eq.~(\ref{eq:complete_gaussian}) shows
that the system undergoes the second-order phase transition with $\beta=1/2$ regardless of the variance $\sigma^2$ of the Gaussian random fields.
In general, we find this phase transition and the same critical exponent for any distribution $G(h)$ that satisfies the condition
\begin{equation}
\int_0^\infty \ h \ G(h) \ v(h)dh > 0.
\label{eq:inequality}
\end{equation}
Otherwise, the random field Kuramoto model undergoes a first-order phase transition.

\subsection{Scale-free networks}
\subsubsection{Homogeneous random fields, degree exponent $\gamma>5$}
Solving Eq. (\ref{eq:expansion}) for scale-free networks with $P(q)=Cq^{-\gamma}$ at $\gamma>5$, we  find the same critical behavior as in the case of the complete graph. In the case of homogeneous random fields, we find the critical coupling,
\begin{equation}
K_c=2\sqrt{h^2+1}\frac{\langle q \rangle}{\langle q^2 \rangle}=2\sqrt{h^2+1}\frac{(\gamma-3)}{(\gamma-2)},
\label{eq:Kc_sf_homogeneous}
\end{equation}
and
\begin{equation}
r=\frac{K_c^2}{2}\sqrt{\frac{2\langle q^2 \rangle}{K^3(2-h^2)\langle q^4 \rangle}}(K-K_c)^{1/2}.
\label{eq:sf5_homogeneous}
\end{equation}
Actually, Eq.~(\ref{eq:Kc_sf_homogeneous}) is valid for any $\gamma>2$.

\subsubsection{Homogeneous random fields, $3<\gamma\leq 5$}
Since $\langle q^4 \rangle$ diverges when $\gamma\leq 5$, one cannot readily use Eq.~(\ref{eq:expansion}) to find the critical behavior. In order to get rid of the divergencies, we integrate by parts twice Eq.~(\ref{eq:r}) before making the Taylor expansion. We find
\begin{eqnarray}
r&=&\frac{C}{\langle q \rangle} \bigg[\frac{Kr}{2\sqrt{h^2+1}(\gamma-3)} + \frac{v(h)(Kr)^3}{6(\gamma -5)}\nonumber\\
& & +\frac{(Kr)^{\gamma-2}}{(2 - \gamma) (3 - \gamma)} \int_0^\infty Y''(x) x^{-\gamma + 3}  dx\bigg] ,
\label{eq:r_parts}
\end{eqnarray}
where we introduced a function,
\begin{equation}
Y(x)=\frac{1}{2\pi} \int_0^{2\pi} \frac{x + h \cos\phi}{\sqrt{(x+h \cos\phi)^2+h^2 \sin^2\phi +1}+1} \ d\phi.
\end{equation}
Note that the third singular term in Eq.~(\ref{eq:r_parts}) is negligible small when $\gamma>5$, but non-negligible when $\gamma\leq5$. One can also see that the case $\gamma=3$ needs further considerations. When $3<\gamma<5$, the first and third terms in Eq.~(\ref{eq:r_parts}) are leading, yielding
\begin{equation}
r=\left[\frac{\langle q \rangle (2-\gamma)K^{2-\gamma}}{\langle q^2 \rangle K_c \int_0^\infty Y''(x) x^{-\gamma+3} dx} \right]^{\frac{1}{\gamma-3}} (K-K_c)^{1/(\gamma-3)}.
\label{eq:r35}
\end{equation}
This solution is real for any $h$, meaning that the transition to the synchronous state is always continuous independently on the magnitude of the local random fields, in striking contrast to scale-free networks with $\gamma>5$. Also, the exponent is non-mean-field, $\beta=1/(\gamma-3)$. The critical coupling is given by Eq.~(\ref{eq:Kc_sf_homogeneous}).

\subsubsection{Homogeneous random fields,  $2<\gamma \leq 3$}
When $2<\gamma < 3$, the second moment $\langle q^2 \rangle$ diverges. In this case, Eq.~(\ref{eq:Kc_sf_homogeneous}) gives the critical coupling $K_c=0$. Furthermore, the third term in Eq.~(\ref{eq:r_parts}) is the leading term, and so the order parameter $r$ is
\begin{equation}
r=\left[\frac{\langle q \rangle(2-\gamma)(3-\gamma)}{C \int_0^\infty Y''(x) x^{-\gamma+3}dx}\right]^\frac{1}{\gamma-3}K^{(2-\gamma)/(\gamma-3)}.
\label{eq:r23}
\end{equation}
Thus, $r>0$ for any $K>0$. The order parameter $r$ follows a non-mean-field scaling law with $\beta=(2-\gamma)/(\gamma-3)$.

Note that both Eqs.~(\ref{eq:r35}) and (\ref{eq:r23}) indicate that $\beta\to\infty$ for $\gamma\to3$, suggesting an infinite-order phase transition at $K_c=0$. Indeed, solving Eq.~(\ref{eq:r}) by use of the same method, at $\gamma=3$ we find
\begin{equation}
r=\frac{2}{K \langle q \rangle} \exp\Bigl(-\frac{4\sqrt{h^2+1}}{\langle q \rangle K}\Bigl)
\end{equation}
at $K>0$.

\begin{table}[t]
\caption{Critical behavior of the order parameter of the Kuramoto model on a complete graph and complex networks with degree distribution $P(q)\propto q^{-\gamma}$ in the presence of homogeneous and Gaussian random fields. In the case of complete graph and scale-free network with $\gamma>5$ (*), the presented critical behavior occurs in the case of Gaussian fields and homogeneous fields at $h<\sqrt{2}$. At $h>\sqrt{2}$ the transition is discontinuous. The function $f(\sigma)$ is defined by Eq.~(\ref{eq:GF-Kc}),  $f(\sigma)=K_c(\sigma) \langle q^2\rangle /  \langle q \rangle$.}
\begin{center}
\begin{tabular}{c | c | c }
\hline
\hline
Network &  \multicolumn{2}{c}{Field}                              \\ \cline{2-3}
              & Homogeneous  & Gaussian                          \\
\hline
\hline
Complete & \multicolumn{2}{c}{$r \propto \sqrt{K-K_c}\quad$(*)}   \\ \cline{2-3}
Graph      &  $K_c=2\sqrt{h^2+1}$     &  $K_c=f(\sigma)$ \\
\hline
\hline
Scale-free       & \multicolumn{2}{c}{$r \propto \sqrt{K-K_c}\quad$(*)}  \\ \cline{2-3}
$\gamma>5$  & $K_c=2\sqrt{h^2+1}\frac{\langle q\rangle}{\langle q^2\rangle}$ & $K_c=f(\sigma)\frac{\langle q\rangle}{\langle q^2\rangle}$ \\
\hline
\hline
$3<\gamma\le5$ & \multicolumn{2}{c}{$r \propto (K-K_c)^{1/(\gamma-3)}$}  \\ \cline{2-3}
                            & $K_c=2\sqrt{h^2+1}\frac{\langle q\rangle}{\langle q^2\rangle}$ & $K_c=f(\sigma)\frac{\langle q\rangle}{\langle q^2\rangle}$ \\
\hline
\hline
$\gamma=3$ & \multicolumn{2}{c}{$r \propto \frac{1}{K}e^{-1/K}$}  \\ \cline{2-3}
                      & \multicolumn{2}{c}{$K_c=0$} \\
\hline
\hline
$2<\gamma<3$ & \multicolumn{2}{c}{$r \propto (K-K_c)^{(2-\gamma)/(\gamma-3)}$}  \\ \cline{2-3}
                          & \multicolumn{2}{c}{$K_c=0$} \\
\hline
\hline
\end{tabular}
\end{center}
\label{table}
\end{table}

\subsubsection{Gaussian random fields, $\gamma >2$}
Now we consider Gaussian fields. Analyzing Eq.~(\ref{eq:expansion}), we find $K_c$ as a function 
the first and second moments of the degree distribution,
\begin{equation}
K_c=\frac{\langle q \rangle}{\langle q^2 \rangle} \frac{1}{\int_0^\infty dh  h  G(h) [2\sqrt{h^2+1}]^{-1}},
\label{eq:GF-Kc}
\end{equation}
where $G(h)$ is given by Eq.~(\ref{eq:gaussian}).
Thus $K_c=0$ at $2<\gamma \leq3$ as in the case of homogeneous random field.

In contrast to the case of homogeneous random field,
there is no first-order phase transition regardless of $\sigma$ and $\gamma$. The critical behavior of the order parameter at $\gamma >2$ is qualitatively the same as in the case of the continuous phase  transition studied above for homogenous random fields [see Table~\ref{table} that summarizes our analytical results].
\subsubsection{Numerical simulations}
\begin{figure}
\includegraphics[width=0.4\textwidth]{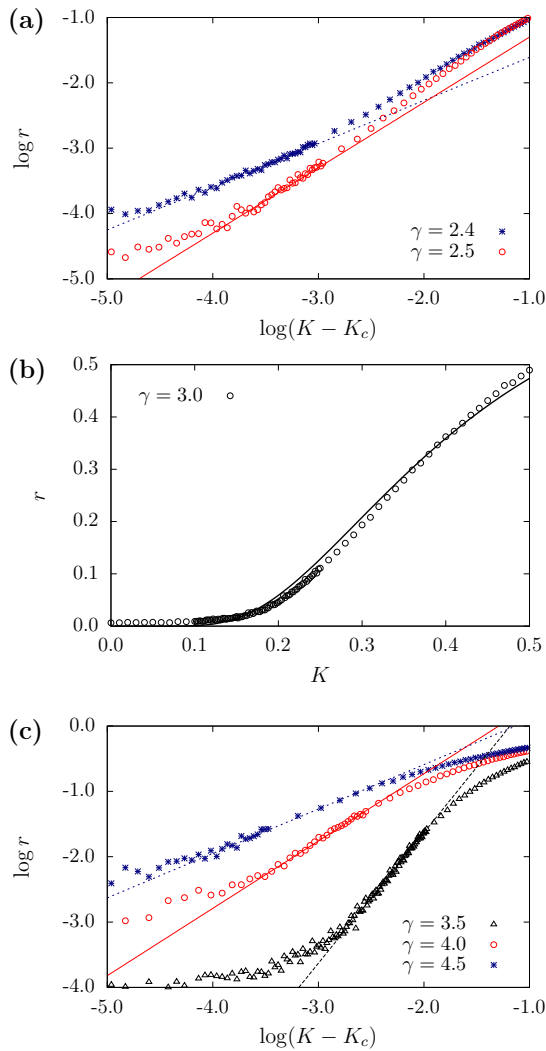}
\caption{Critical behavior of the order parameter in scale-free networks in the presence of homogeneous random fields ($h=2$). The symbols correspond to simulations of the model (average over $30$ network realizations), whereas the lines in panels (a) and (c) are the linear fits. The line in panel (b) is $f(K)\propto \frac{1}{K}\exp(-1/K)$. The networks were generated using the static model ($N=2\times10^4$ and $\langle q \rangle=10$) \cite{Goh_2001,Lee_2004}. The differential equations for each oscillator were solved using a fourth-order Runge-Kutta method with time step $\Delta t=0.01$.
\label{fig:sim_sf}}
\end{figure}

We also performed simulations of the random field Kuramoto model in the case of homogeneous random fields. Figure~\ref{fig:sim_sf} shows the critical behavior of the order parameter in scale-free networks in three cases:
(a) $2<\gamma<3$, (b) $\gamma=3$, and (c) $3<\gamma<5$.
Table~\ref{table2} compares $K_c$ and the critical exponent $\beta$ between simulations and our analytical results. Each linear regression was obtained for such $K_c$ that maximize the linear correlation coefficient. One can see that our analytical results are in good agreement with the simulations. The difference between the numerical results and the theory, when $K$  is close to $K_c$, is caused by finite-size fluctuations that smear the phase transition.

\begin{table}[t]
\caption{Comparison of the critical coupling $K_c$ and the order parameter exponent $\beta$ between the simulations (S) in Fig.~\ref{fig:sim_sf} and the analytical (A) results in Table~\ref{table} for random field magnitude $h=2$. }
\begin{center}
\begin{tabular}{c | c | c |c | c }
\hline
\hline
$\gamma$ & $\beta$ (A) & $\beta$ (S) & $K_c$ (A) & $K_c$ (S) \\
\hline
$2.4$ & $\frac{(2-\gamma)}{(\gamma-3)}\approx0.667$ & $0.66\pm0.01$ & $0$ & $0.052$ \\
\hline
$2.5$ & $\frac{(2-\gamma)}{(\gamma-3)}=1$ & $1.00\pm0.03$ & $0$ & $0.048$ \\
\hline
$3.5$ & $\frac{1}{(\gamma-3)}=2$ & $2.00\pm0.02$ & $2\sqrt{1+h^2}\frac{\langle q \rangle}{ \langle q^2 \rangle}\approx 0.266$ & $0.211$ \\
\hline
$4.0$ & $\frac{1}{(\gamma-3)}=1$ & $1.03\pm0.02$ & $2\sqrt{1+h^2}\frac{\langle q \rangle}{ \langle q^2 \rangle}\approx 0.319$ & $0.322$ \\
\hline
$4.5$ & $\frac{1}{(\gamma-3)}\approx0.667$ & $0.68\pm0.02$ & $2\sqrt{1+h^2}\frac{\langle q \rangle}{ \langle q^2 \rangle}\approx 0.349$ & $0.369$ \\
\hline
\hline
\end{tabular}
\end{center}
\label{table2}
\end{table}


\section{Conclusion}
\label{conclusion}

In this paper, we studied the impact of random pinning fields on the synchronization of phase oscillators in the Kuramoto model on a complete graph and uncorrelated complex networks with different degree distributions (scale-free networks).
We considered random pinning fields whose directions are uniformly distributed and the field magnitudes are either homogeneous (i.e., they are equal for all oscillators) or heterogeneous (Gaussian). First, we found that in the case of homogeneous random fields, in the fully connected network and scale-free networks with the degree exponent $\gamma>5$, there is a critical random field magnitude above which the second-order phase transition gives place to a first-order phase transition.
However, in contrast to networks with rapidly decaying degree distributions, in scale-free networks with $3 < \gamma \leq 5$, the phase transition remains of second-order at any random field magnitude though the critical coupling strongly depends on it. Furthermore, we showed that if $2 < \gamma \leq 3$ then the synchronization transition also remains continuous and occurs at zero coupling $K_c=0$ independently on the field distribution. In the case $\gamma=3$  the synchronization transition is of the infinite order.

Second, we demonstrated that in the Kuramoto model with heterogeneous random fields even
strong Gaussian random fields do not suppress the synchronization though the critical coupling depends strongly on the field variance.
The continuous phase transition into the synchronized state is characterized by the same critical exponents as the synchronization transition in the absence of the fields.
The critical behavior of the order parameter and the
critical coupling are summarized in Table~\ref{table}.
We also carried out simulations of the Kuramoto model on complete graphs and scale-free networks in the presence of homogenous random fields. These simulations confirmed our analytical results.


Interestingly, the critical behaviour of the order parameter, relaxation rate, and susceptibility of the Kuramoto model on uncorrelated random complex networks  \cite{Yoon_2015} is characterized by the same critical exponents as the Ising model. However, the two models no longer present the same critical behaviour in the presence of random fields in both complete graph and scale-free networks.
Whereas Gaussian random fields in the Ising model suppress a phase transition when the random field variance $\sigma$ is above a critical value $\sigma_c$ \cite{dhar1997zero,Dorogovtsev_2008,Dorogovtsev_2008arxiv,Lee_2006}, in the Kuramoto model a sufficiently strong coupling can always prevail over the random fields and results in synchronization of oscillators for any $\sigma$ at $\gamma >2$. In the case of scale-free networks, Gaussian fields elicit different behaviors in the two models: while there is a first-order phase transition in the Ising model, the transition is of second-order in the Kuramoto model.

\section{Acknowledgements}
This work was partially supported by FET IP Project MULTIPLEX 317532,
A.V.G. and S.Y. are grateful to LA I3N for the grant PEST UID/CTM/50025/2013.
M.A.L. acknowledges the financial support of the Medical Research Council (MRC) via Programme Grant MR/K013998/01.

\bibliography{mybib}
\end{document}